\documentclass[12pt]{iopart}
\usepackage{graphicx}
\usepackage{color}
\newcommand{\ket}[1]{\left|{}#1 \right>}

\newcommand{\interproduct}[2]{\langle {}#1 | {}#2 \rangle}

\begin{document}

\title{Unified approach to the nonlinear Rabi models}

\author{Liwei Duan}

\address{Department of Physics, Zhejiang Normal University, Jinhua 321004, China}
\ead{duanlw@gmail.com}
\vspace{10pt}

\begin{abstract}
An analytical approach is proposed to study the two-photon, two-mode and intensity-dependent Rabi models. By virtue of the su(1,1) Lie algebra, all of them can be unified to the same Hamiltonian with $\mathcal{Z}_2$ symmetry. There exist exact isolated solutions, which are located at the level crossings between different parities and  correspond to eigenstates with finite dimension. Beyond the exact isolated solutions, the regular spectrum can be achieved by finding the roots of the G-function. The corresponding eigenstates are of infinite dimension. It is noteworthy that the expansion coefficients of the eigenstates present an exponential decay behavior. The decay rate decreases with increasing coupling strength.  When the coupling strength tends to the spectral collapse point $g \rightarrow \omega / 2$, the decay rate tends to zero which prevents the convergence of the wave functions. This work paves a way for the analysis of novel physics in nonlinear quantum optics.
\end{abstract}

%
\vspace{2pc}
\noindent{\it Keywords}: nonlinear Rabi model, su(1,1) Lie algebra, exact solution

\submitto{\NJP}
%
%
%

\section{Introduction}

As a paradigmatic model to study the light-matter interacting systems, the Rabi model has been proposed for more than $80$ years \cite{PhysRev.51.652,scully_zubairy_1997,Braak_2016}. Renewed attention has been paid to the quantum Rabi model over the last decades, due to the burst of experiments that push into ultrastrong and even deep strong coupling regimes \cite{RevModPhys.91.025005,frisk_kockum_ultrastrong_2019,RevModPhys.90.031002}, the emergence of the quantum phase transition in the finite component systems \cite{PhysRevLett.115.180404,PhysRevLett.117.123602,PhysRevLett.119.220601,Cai2021,PhysRevLett.127.063602}, as well as the breakthrough of the analytical exact solutions obtained from the G-functions in the  Bargmann space \cite{PhysRevLett.107.100401} and Bogoliubov operator approach \cite{PhysRevA.86.023822}. The quantum Rabi model serves as a building block for the quantum information processing \cite{RevModPhys.91.025005}, and forms a connecting link between mathematics, physics, and technology \cite{Braak_2016}. The quantum Rabi model originally describes a two-level system linearly interacting with a single bosonic mode \cite{1443594}. Recently, a generalized Rabi model has stepped into the spotlight which considers the nonlinear interaction between the two-level system and the bosonic field. Among them, two-photon, two-mode and intensity-dependent Rabi models are three typical ones which introduce different forms of nonlinear interactions.

The two-photon and two-mode Rabi models describe the transitions of the two-level system accompanied by emitting or absorbing two photons in single- and two-mode bosonic field respectively. They can be used to describe the second-order process with consequently small coupling strengths in different physical setups \cite{PhysRevA.92.033817}. Two-photon processes can generate high-order correlations between the emitted photons which is of great significance in quantum optics and quantum information science \cite{PhysRevA.45.4951,PhysRevB.81.035302}. Recently, implementations of the two-photon Rabi models in the trapped ion \cite{PhysRevA.92.033817,PhysRevA.95.063844} and superconducting circuits \cite{PhysRevA.97.013851,PhysRevA.98.053859} have been proposed which can reach the ultrastrong coupling regime. The increase in the coupling strength prompts us to search for more accurate methods beyond the rotating wave approximation (RWA) \cite{L__2017,PhysRevA.99.013815,PhysRevA.101.033827}. Based on the numerical diagonalization in a truncated basis, Ng \etal found that there exist significant differences in the energy spectra with and without RWA \cite{NG2000463}.  Emary and Bishop found  exact isolated solutions for the two-photon Rabi model based on the Bogoliubov transformations \cite{Emary_2002}. Furthermore, Chen \etal first proposed a G-function based on the Bogoliubov operator approach, with which they achieved the exact isolated solutions and the complete regular spectrum of the two-photon and two-mode Rabi models \cite{PhysRevA.86.023822,Duan_2016,Duan_2015}. Braak provided a rigorous proof of validity of Chen's G-function based on the normalizability of the wavefunctions in the Bargmann space \cite{Braak_2022}.

Pioneered by Buck and Sukumar, they proposed an intensity-dependent Jaynes-Cummings model, namely the Buck-Sukumar model, to study the collapse and revival behavior of the two-level system \cite{BUCK1981132}. The intensity-dependent Rabi model can be regarded as a generalization of the Buck-Sukumar model which introduces the counter-rotating wave terms and the Holstein-Primakoff realization of the su(1,1) operators \cite{NG2000463,PhysRevA.93.043814,Penna_2017}. The trapped ion far away from the Lamb-Dicke regime can be used to simulate the nonlinear Rabi model \cite{PhysRevA.52.4214}, and it can be used to generate arbitrary $n$-phonon Fock states \cite{PhysRevA.97.023624}. 
To the best of our knowledge, neither the exact isolated solutions nor the regular spectrum have been found in this model.

Although the Hamiltonians of the two-photon, two-mode and intensity-dependent Rabi models are quite different, they share some common features: (i) One can introduce the su(1,1) Lie algebra to describe the bosonic parts of three Hamiltonians. The Bargmann index can be used to characterize different Hilbert subspaces. (ii) All of them exist spectral collapse phenomena \cite{NG2000463,PhysRevA.92.033817,Duan_2016,Penna_2017}. When the coupling strength is large enough, the discrete energy levels tend to form a continuous energy band except for some low-lying states \cite{Duan_2016}. Beyond the spectral collapse point, the nonlinear Rabi models become no longer self-adjoint.
In this paper, we employ the su(1,1) Lie algebra to unify three models to a general Hamiltonians with $\mathcal{Z}_2$ symmetry. Then, the analytical solutions to the general Hamiltonian are achieved by employing the Bogoliubov operator approach. 

The paper is structured as follows. In section \ref{sec:su11}, we revisit the su(1,1) Lie algebra. In section \ref{sec:Rabi}, we introduce a general Hamiltonian which recovers three nonlinear Rabi models by employing different realizations of su(1,1) algebra. In section \ref{sec:method}, we construct an ansatz for the general Hamiltonian by choosing appropriate basis states. The asymptotic behavior of the expansion coefficients of the ansatz is analyzed. The condition to achieve exact isolated solutions is given. Beyond the exact isolated solutions, the regular spectrum is achieved by solving the G-function. The energy spectrum and the behavior of the eigenstates can be found in section \ref{sec:result}. Finally, a brief summary is given in section \ref{sec:summary}.

\section{SU(1,1) group} \label{sec:su11}
The group theory has been employed in various branches in quantum optics \cite{Wodkiewicz:85,Gerry:01}.
We begin by briefly reviewing the basic properties of the SU(1,1) group and its associate su(1,1) algebra.
The SU(1,1) group is non-compact. The generators associated with SU(1,1) group satisfy
\begin{eqnarray}
	\left[\hat{K}_0, \hat{K}_{\pm}\right] = \pm \hat{K}_{\pm}, \quad \left[\hat{K}_+, \hat{K}_-\right] = -2 \hat{K}_0 .
\end{eqnarray}
The corresponding Casimir $\hat{C}$ operator can be written as
\begin{eqnarray}
	\hat{C} &=& \hat{K}_0^2 - \frac{1}{2} \left(\hat{K}_+ \hat{K}_- + \hat{K}_- \hat{K}_+\right) ,
\end{eqnarray}
which commutes with all the elements of the su(1,1) Lie algebra. One can choose the basis state $\ket{k,m}$, which satisfies the following relations,
\numparts
	\begin{eqnarray}
		\hat{K}_0 \ket{k,m} &=& (k + m) \ket{k,m}, \\
		\hat{K}_+ \ket{k,m} &=& \sqrt{(m + 1)(m + 2k)} \ket{k, m + 1}, \\
		\hat{K}_- \ket{k,m} &=& \sqrt{m(m + 2k - 1)} \ket{k, m - 1}, \\
		\hat{C} \ket{k,m} &=& k (k - 1) \ket{k, m} ,
	\end{eqnarray}
\endnumparts
with $m = 0, 1, 2, \dots$ All states can be obtained from the lowest one $\ket{k, 0}$ by successive actions of the raising operator $\hat{K}_+$ according to 
\begin{eqnarray}
	\ket{k, m} = \sqrt{\frac{\Gamma(2 k)}{m! \Gamma(2 k + m)}} \hat{K}_+^m \ket{k, 0} .
\end{eqnarray}
The number $k$ is known as the Bargmann index which separates different irreducible representations.

\section{Nonlinear Rabi model} \label{sec:Rabi}

A general nonlinear Rabi model with an su(1,1) coupling scheme can be written as 
\begin{eqnarray}
	\hat{H} = \frac{\epsilon}{2} \hat{\sigma}_z + \omega \hat{K}_0 + g \hat{\sigma}_x \left(\hat{K}_+ + \hat{K}_-\right) , \label{eq:H}
\end{eqnarray}
where $\epsilon$ and $\omega$  correspond to the frequency of the two-level system and bosonic field respectively, $g$ is the coupling strength. Like the linear Rabi model, the nonlinear one has $\mathcal{Z}_2$ symmetry. The parity operator can be defined as $\hat{\Pi} 
= -\hat{\sigma}_z \otimes \hat{T}$ with $\hat{T} = \exp \left[\rmi \pi \left(\hat{K}_0 - k\right)\right]$.
We can easily verify that
\begin{eqnarray}
	\hat{\Pi} \hat{\sigma}_z \hat{\Pi}^{\dagger} = \hat{\sigma}_z, \quad \hat{\Pi} \hat{\sigma}_x \hat{\Pi}^{\dagger} = - \hat{\sigma}_x, \quad \hat{\Pi} \hat{K}_0 \hat{\Pi}^{\dagger} = \hat{K}_0, \quad \hat{\Pi} \hat{K}_{\pm} \hat{\Pi}^{\dagger} = -\hat{K}_{\pm} , \nonumber
\end{eqnarray} 
which leads to $\hat{\Pi} \hat{H} \hat{\Pi}^{\dagger} = \hat{H}$ and $\left[\hat{H}, \hat{\Pi}\right] = 0$.
The parity operator $\hat{\Pi}$ has eigenvalues $\Pi = \pm 1$, and it can separate the whole Hilbert space into two subspaces with even and odd parities respectively. Unlike the linear Rabi model, the nonlinear one also commutes with the Casimir operator $\hat{C}$, which separates the whole Hilbert space into different subspaces indexed by the Bargmann index $k$.

Such a Hamiltonian has been studied by Penna \etal \cite{Penna_2017} who mainly focused on the two-mode and Holstein-Primakoff realizations of the su(1,1) algebra. Depending on the choice of the realizations, $\hat{H}$ can be expressed in different forms.

\subsection{Two-photon Rabi model}
In the one-mode bosonic realization, the generators can be expressed as
	\begin{eqnarray}\label{eq:one-mode}
		\hat{K}_0 = \frac{1}{2} \left(\hat{a}^{\dagger} \hat{a} + \frac{1}{2}\right), \quad
		\hat{K}_+ = \frac{1}{2} \left(\hat{a}^{\dagger}\right)^2, \quad
		\hat{K}_- = \frac{1}{2} \hat{a}^2 ,
	\end{eqnarray}
where $\hat{a}$ ($\hat{a}^{\dagger}$) is the bosonic annihilation (creation) operator. The corresponding Bargmann index is $k = \frac{1}{4}$ or $\frac{3}{4}$. Given the Fock states $\ket{n}_{a}$ which satisfies $\hat{a}^{\dagger} \hat{a} \ket{n}_{a} = n \ket{n}_{a}$, the basis state $\ket{k, m}$ can be rewritten as
\begin{eqnarray}
	\ket{k, m} = \ket{2 \left(m + k - \frac{1}{4}\right) }_a. 
\end{eqnarray}
Therefore, the number of bosons is even and odd for $k = \frac{1}{4}$ and $\frac{3}{4}$ respectively.

One can obtain the well-known two-photon Rabi model \cite{NG2000463,Emary_2002,Duan_2016} by substituting $\hat{K}_0$ and $\hat{K}_{\pm}$ in (\ref{eq:H}) with those in (\ref{eq:one-mode}), which leads to
\begin{eqnarray} \label{eq:H_2P}
	\hat{H}_{\mathrm{2p}} &=& \hat{H} - \frac{\omega_{\mathrm{2p}}}{2}, \nonumber\\
	&=& \frac{\epsilon}{2} \hat{\sigma}_z + \omega_{\mathrm{2p}} \hat{a}^{\dagger} \hat{a} + g_{\mathrm{2p}} \hat{\sigma}_x \left(\left(\hat{a}^{\dagger}\right)^2 + \hat{a}^2\right) ,
\end{eqnarray}
with $\omega_{\mathrm{2p}} = \omega / 2$, $g_{\mathrm{2p}} = g / 2$. 

\subsection{Two-mode Rabi model}
In the two-mode bosonic realization, the generators can be expressed as
	\begin{eqnarray} \label{eq:two-mode}
		\hat{K}_0 = \frac{1}{2} \left(\hat{a}^{\dagger} \hat{a} + \hat{b}^{\dagger} \hat{b} + 1\right), \quad \hat{K}_+ = \hat{a}^{\dagger} \hat{b}^{\dagger}, \quad \hat{K}_- = \hat{a} \hat{b} .
	\end{eqnarray}
The corresponding Bargmann index is $k = \frac{1}{2},~1,~\frac{3}{2},~\dots $ Given the Fock states $\ket{n}_{s}$ ($s=a, b$) which satisfies $\hat{s}^{\dagger} \hat{s} \ket{n}_{s} = n \ket{n}_{s}$, the basis state $\ket{k, m}$ can be rewritten as
\begin{eqnarray}
	\ket{k, m} = \ket{m + 2 k - 1}_a \otimes \ket{m}_b.
\end{eqnarray}
Therefore, the Bargmann index are related with the number difference between two modes.

One can obtain the two-mode Rabi model \cite{Duan_2015,Penna_2017,doi:10.1063/1.4826356} by substituting $\hat{K}_0$ and $\hat{K}_{\pm}$ in (\ref{eq:H}) with those in (\ref{eq:two-mode}), which leads to
\begin{eqnarray}\label{eq:H_2M}
	\hat{H}_{\mathrm{2m}} &=& \hat{H} - \omega_{\mathrm{2m}} \\
	&=& \frac{\epsilon}{2} \hat{\sigma}_z + \omega_{\mathrm{2m}} \left(\hat{a}^{\dagger} \hat{a} + \hat{b}^{\dagger} \hat{b}\right) + g_{\mathrm{2m}} \hat{\sigma}_x \left(\hat{a}^{\dagger} \hat{b}^{\dagger} + \hat{a} \hat{b}\right) , \nonumber
\end{eqnarray}
with $\omega_{\mathrm{2m}} = \omega / 2$ and $g_{\mathrm{2m}} = g$.

\subsection{Intensity-dependent Rabi model}
In the Holstein-Primakoff realization, the generators can be expressed as
	\begin{eqnarray} \label{eq:HP}
		\hat{K}_0 = \hat{a}^{\dagger} \hat{a} + k , \quad \hat{K}_+ = \sqrt{\hat{a}^{\dagger} \hat{a} + 2k - 1} \hat{a}^{\dagger} , \quad \hat{K}_- = \hat{a} \sqrt{\hat{a}^{\dagger} \hat{a} + 2k - 1} ,
	\end{eqnarray}
where $k$ is the Bargmann index. In this case, the basis state $\ket{k, m}$ is nothing but the Fock state, namely, $\ket{k, m} = \ket{m}_a$.

One can obtain the intensity-dependent Rabi model \cite{Penna_2017,Rodriguez-Lara:14} by substituting $\hat{K}_0$ and $\hat{K}_{\pm}$ in (\ref{eq:H}) with those in (\ref{eq:HP}), which leads to
\begin{eqnarray}\label{eq:H_I}
	\hat{H}_{\mathrm{I}} &=& \hat{H} - k \omega_{\mathrm{I}} \\
	&=& \frac{\epsilon}{2} \hat{\sigma}_z + \omega_{\mathrm{I}} \hat{a}^{\dagger} \hat{a} + g_{\mathrm{I}} \hat{\sigma}_x \left(\sqrt{\hat{a}^{\dagger} \hat{a} + 2k - 1} \hat{a}^{\dagger} + \hat{a} \sqrt{\hat{a}^{\dagger} \hat{a} + 2k - 1}\right) ,\nonumber
\end{eqnarray}
with $\omega_{\mathrm{I}} = \omega$ and $g_{\mathrm{I}} = g$. The well-known Buck-Sukumar model \cite{BUCK1981132} is recovered  if we choose $k = \frac{1}{2}$ and perform the RWA.

\section{Methods} \label{sec:method}

Up to a global shift, three nonlinear Rabi models can be expressed in a united form $\hat{H}$ (\ref{eq:H}) in terms of the su(1,1) generators. Instead of solving three nonlinear Rabi models individually, we only need to find  the analytical solutions to $\hat{H}$ itself.

In the basis state of $\{\ket{\pm}\}$ which satisfy $\hat{\sigma}_x \ket{\pm} = \pm \ket{\pm}$, Hamiltonian (\ref{eq:H}) can be written in a matrix form,
\begin{equation}
	\hat{H} = \left(
	\begin{array}{cc}
		\omega \hat{K}_0 + g \left(\hat{K}_+ + \hat{K}_-\right) & -\frac{\epsilon}{2}\\
		-\frac{\epsilon}{2} & \omega \hat{K}_0 - g \left(\hat{K}_+ + \hat{K}_-\right)
	\end{array}
	\right) . \label{eq:H2}
\end{equation}

Since diagonal elements only  consist of the linear combination of the su(1,1) generators, one can easily achieve its eigenstates and eigenvalues which satisfy
\begin{equation}
	\left[\omega \hat{K}_0 \pm g \left(\hat{K}_+ + \hat{K}_-\right)\right] \ket{k,m}_{\pm} = \beta \left(k + m\right) \ket{k,m}_{\pm} ,
\end{equation}
with $\beta = \sqrt{\omega^2 - 4 g^2}$, $r = \mathrm{arctanh} \left(\frac{2g}{\omega}\right)$ and
\begin{eqnarray}
	\ket{k,m}_{\pm} &=& \hat{S} (\mp r) \ket{k, m},\\
	\hat{S} (\mp r) &=& \exp \left[\mp \frac{r}{2} \left(\hat{K}_+ - \hat{K}_-\right)\right]  .
\end{eqnarray}
For the one-mode and two-mode realizations, $\hat{S} (\mp r)$ corresponds to the well-known Bogoliubov transformation or squeezing operator \cite{scully_zubairy_1997}. It should be noted that when $g \rightarrow \omega / 2$, $\beta$ tends to zero, which leads to spectral collapse \cite{PhysRevA.92.033817,NG2000463,Duan_2016,Braak_2022,Penna_2017}.  Beyond the spectral collapse point ($g>\omega / 2$), the nonlinear Rabi model becomes no longer self-adjoint \cite{Braak_2022}. Therefore, we only focus on $0< g < \omega / 2$ in this paper.

To achieve the eigenstates of $\hat{H}$ (\ref{eq:H2}), one can construct an ansatz which is written as a superpostion of $\{\ket{k, m}_+\}$, namely,
\begin{eqnarray} \label{eq:psi}
	\ket{\psi} = \sum_{m = 0}^{+ \infty} \left(
	\begin{array}{c}
		c_m \ket{k, m}_+\\
		d_m \ket{k, m}_+
	\end{array}
	\right). 
\end{eqnarray}
From the Schr\"{o}dinger equation $\hat{H} \ket{\psi} = E \ket{\psi}$, the expansion coefficients $c_m$ and $d_m$ should satisfy
\begin{eqnarray}
	\beta \left(k + m\right) c_m - \frac{\epsilon}{2} d_m = E c_m, \label{eq:up}
\end{eqnarray}
	\begin{eqnarray}\fl
		- \frac{\epsilon}{2} c_m + \beta \left[\cosh 2r \left(k + m\right)  d_m - \frac{\sinh 2r}{2} \left(\sqrt{m(m + 2k - 1)} d_{m - 1} \right.\right.\nonumber\\
		\left.\left.+ \sqrt{(m + 1)(m + 2k)} d_{m + 1}\right)\right] = E d_m. \label{eq:down}
	\end{eqnarray}
Equation (\ref{eq:up}) leads to 
\begin{eqnarray}
	c_m = \frac{\epsilon / 2}{\beta \left(k + m\right) - E} d_m . \label{eq:c_m}
\end{eqnarray}  
Substituting the above equation into (\ref{eq:down}), we can obtain a three-term recurrence relation for $d_m$, namely,
\begin{eqnarray}
	d_{m + 1} &=& T_m d_m - R_{m - 1} d_{m - 1}  . \label{eq:3recur}
\end{eqnarray}
with 
\begin{eqnarray}
	T_m &=& \frac{2\left(\beta \cosh 2r \left(k + m\right)  - \frac{\epsilon^2/4}{\beta \left(k + m\right) - E} - E\right)}{\beta \sinh 2r  \sqrt{(m + 1)(m + 2k)}}, \\
	R_m &=& \sqrt{\frac{(m + 1)(m + 2k)}{(m + 2)(m + 2k + 1)}}.
\end{eqnarray}
In the limit of $m \rightarrow + \infty$,  we find that
\begin{eqnarray}
	\lim_{m \rightarrow +\infty} \frac{d_{m + 1}}{d_m} = \left\{\begin{array}{c}
		\tanh r  = \frac{2 g}{\omega} < 1\\
		\\
		\coth r = \frac{\omega}{2 g} > 1
	\end{array}\right. .  \label{eq:limit}
\end{eqnarray}
Therefore, equation (\ref{eq:3recur}) has two linearly independent solutions $d_{m}^{(0)}$ and $d_{m}^{(1)}$ with the following limit behaviors respectively,
\begin{eqnarray}
	\lim_{m \rightarrow +\infty} \frac{d_{m+1}^{(0)}}{d_m^{(0)}} = \frac{2 g}{\omega}, \quad \lim_{m \rightarrow +\infty} \frac{d_{m+1}^{(1)}}{d_m^{(1)}} = \frac{\omega}{2 g}.
\end{eqnarray}
Generally, $d_m$ can be written as
\begin{eqnarray} \label{eq:general_d_m}
	d_m = A^{(0)} d_m^{(0)} + A^{(1)} d_m^{(1)} .
\end{eqnarray}
Due to $0<g<\omega / 2$, we expect that the eigenstates correspond to $A^{(1)} = 0$, otherwise it will lead to divergence.

\subsection{Exact isolated solutions}

Special attention should be paid to (\ref{eq:up}). When the energy satisfies $E = \beta \left(k + M\right)$ with $M=1,2,3,\dots$, the expansion coefficient $c_{m=M}$ and the corresponding wavefunction will not diverge only if $d_{m=M} = 0$.
	$E = \beta \left(k + M\right)$ is called the baseline energy \cite{doi:10.1063/1.434971,Reik_1982}. 
From the three-term recurrence relation  (\ref{eq:3recur}), $d_M=0$ corresponds to 
\begin{eqnarray} \label{eq:Judd}
	\left|
	\begin{array}{ccccc}
		-T_0 & 1 \\
		R_0 & -T_1 & 1 \\
		& R_1 & -T_2 & 1\\
		&& \ddots & \ddots & \ddots\\
		&&& R_{M - 2} &-T_{M - 1}
	\end{array}
	\right| = 0 ,
\end{eqnarray}
which gives a relation between the system parameters $\epsilon$, $\omega$ and $g$. When this relation is satisfied, $E = \beta \left(k + M\right)$ is the eigenenergy, also known as the exact isolated solutions. The eigenstates corresponding to the exact isolated solutions can be reduced to a closed form as follows,
\begin{eqnarray} \label{eq:psi_EIS}
	\ket{\psi} = \left(
	\begin{array}{c}
		\sum_{m = 0}^{M} c_m \ket{k, m}_+\\
		\sum_{m = 0}^{M - 1} d_m \ket{k, m}_+
	\end{array}
	\right) ,
\end{eqnarray}
where $d_{m<M}$ and $c_{m<M}$ are determined by (\ref{eq:3recur}) and (\ref{eq:c_m}) respectively. $c_{M}$ is determined by (\ref{eq:down}) with $d_M=d_{M+1}=0$, namely,
\begin{eqnarray}
	- \frac{\epsilon}{2} c_M - \beta \frac{\sinh 2r}{2} \sqrt{M(M + 2k - 1)} d_{M - 1} = 0. 
\end{eqnarray}

Such kinds of exact isolated solutions were first discovered by Judd in the Jahn-Teller model \cite{doi:10.1063/1.434971}, which correspond to the level crossings in the energy spectrum. Soon after that, Reik \etal found them in the linear Rabi model \cite{Reik_1982}. The exact isolated solutions for two-photon and two-mode Rabi models were also brought into the spotlight \cite{Emary_2002,Duan_2015,Duan_2016,Braak_2022}. A new type of exact isolated solutions, also known as the dark-like state, were found when generalizing it to the multi-qubit cases \cite{Peng_2014,Peng_2017,PhysRevLett.127.043604}. However, less attention has been paid to those in the intensity-dependent Rabi model.

\begin{table}
	\caption{\label{tab:Juddian} Exact isolated solutions for $k =\frac{1}{4}$ and $\frac{1}{2}$ with $\epsilon=\omega=1$. The energy is located at the baseline  $E = \beta \left(k + M\right)$, while the corresponding coupling strength $g$ is determined by (\ref{eq:Judd}).}
	\begin{indented}
		\lineup
		\item[]\begin{tabular}{@{}lllll}
			\br
			&\centre{2}{$k=\frac{1}{4}$}&\centre{2}{$k=\frac{1}{2}$}\\ \ns&\crule{2}&\crule{2}\\
			& $g$ & $E$ & $g$ & $E$\\
			\mr
			$M = 1$ & 0.353 553 390 6 & 0.883 883 476 5 & 0.306 186 217 8 & 1.185 854 122 6\\
			\mr
			$M = 2$ & 0.220 400 240 2 & 2.019 611 501 3 & 0.199 407 656 4 & 2.292 578 169 8\\
			$M = 2$ & 0.454 731 653 8 & 0.935 514 425 9 & 0.429 179 356 3 & 1.282 625 043 5\\
			\mr
			$M = 3$ & 0.156 833 678 1 & 3.085 982 030 1 & 0.145 778 939 2 & 3.347 936 161 9\\
			$M = 3$ & 0.362 621 090 4 & 2.237 605 006 9 & 0.341 905 645 5 & 2.553 806 169 2\\
			$M = 3$ & 0.478 267 278 3 & 0.947 761 754 5 & 0.463 652 920 3 & 1.310 065 840 3\\
			\br
		\end{tabular}
	\end{indented}
\end{table}
Table \ref{tab:Juddian} gives the exact isolated solutions for $k = \frac{1}{4}$ and $\frac{1}{2}$, where we have fixed $\epsilon = \omega = 1$. The coupling strength $g$ corresponding to $E=\beta (k + M)$ is determined by (\ref{eq:Judd}). When $k=\frac{1}{4}$, it corresponds to the two-photon Rabi model. We exactly reproduce the exact isolated solutions in the two-photon Rabi model presented in  \cite{Emary_2002}. The results for $k=\frac{1}{2}$ is associated with the intensity-dependent Rabi model or the Buck-Sukumar model, which is first discovered to the best of our knowledge.

\subsection{G-functions}

Beyond the exact isolated solutions, one can set $d_0 = 1$ in general, while $d_{m > 0}$ as a function of $E$ can be achieved from the three-term recurrence relation  (\ref{eq:3recur}) successively. Due to the parity symmetry, the eigenstates of $\hat{H}$ should also be those of the parity operator $\hat{\Pi}$, namely, $\hat{\Pi} \ket{\psi} = \Pi \ket{\psi}$. The left-hand side corresponds to
\begin{eqnarray}
	\hat{\Pi} \ket{\psi} &=& -\hat{\sigma}_z \otimes \hat{T} \sum_{m = 0}^{+ \infty} \left(
	\begin{array}{c}
		c_m \ket{k, m}_+\\
		d_m \ket{k, m}_+
	\end{array}
	\right) = \sum_{m = 0}^{+ \infty} \left(
	\begin{array}{c}
		d_m (-1)^m \ket{k, m}_-\\
		c_m (-1)^m \ket{k, m}_-
	\end{array}
	\right), 
\end{eqnarray}
while the right-hand side corresponds to
\begin{eqnarray}
	\Pi \ket{\psi} &=& \Pi \sum_{m = 0}^{+ \infty} \left(
	\begin{array}{c}
		c_m \ket{k, m}_+\\
		d_m \ket{k, m}_+
	\end{array}
	\right) .
\end{eqnarray}
Therefore, 
	\begin{eqnarray}
		\sum_{m = 0}^{+ \infty} d_m (-1)^m \ket{k, m}_- &=& \Pi \sum_{m = 0}^{+ \infty} c_m \ket{k, m}_+ , \label{eq:Pi_up}\\
		\sum_{m = 0}^{+ \infty} c_m (-1)^m \ket{k, m}_- &=& \Pi \sum_{m = 0}^{+ \infty} d_m \ket{k, m}_+ .\label{eq:Pi_down}
	\end{eqnarray}
It should be noted that (\ref{eq:Pi_up}) and  (\ref{eq:Pi_down}) are equivalent, since they can be transformed to each other by the unitary transformation $\hat{T}$. Note that \cite{Wodkiewicz:85,Gerry:01}
\begin{eqnarray} \label{eq:overlap}
	\interproduct{k, 0}{k, m}_{+} &=& (-1)^m \interproduct{k, 0}{k, m}_{-} \\
	&=& \left(1 - \xi^2\right)^k \sqrt{\frac{\Gamma(2k + m)}{m! \Gamma(2 k)}} \xi^m , \nonumber
\end{eqnarray}
with $\xi = \tanh \frac{r}{2}$.
Projecting (\ref{eq:Pi_up}) onto $\ket{k, 0}$, we achieve the G-function as follows,
\begin{eqnarray} \label{eq:Gfun}
	G_k^{\Pi}(E) &=& \sum_{m = 0}^{+\infty} d_m (-1)^m \interproduct{k, 0}{k, m}_- - \Pi c_m \interproduct{k, 0}{k, m}_+ \nonumber\\
	&=& \sum_{m = 0}^{+\infty} \Xi_{k,m}^{\Pi}(E) \xi^m ,
\end{eqnarray}
with 
\begin{eqnarray}
	\Xi_{k,m}^{\Pi}(E) = \left(d_m - \Pi c_m \right) \left(1 - \xi^2\right)^k \sqrt{\frac{\Gamma(2k + m)}{m! \Gamma(2 k)}} ,
\end{eqnarray}
where we have employed (\ref{eq:overlap}). $G_k^{\Pi}(E)$ can be regarded as a power series in $\xi$. From (\ref{eq:limit}), we can find that 
\begin{eqnarray}
	\lim_{m \rightarrow +\infty} \frac{\Xi_{k,m+1}^{\Pi}}{\Xi_{k,m}^{\Pi}} 
	&\le& \coth r, 
\end{eqnarray}
which indicates that the radius of convergence satisfies $R \ge \tanh r$. Therefore, $G_k^{\Pi}(E)$ is well-defined and will always converge due to $\xi < R$. The roots of $G_k^{\Pi}(E) = 0$ determine the eigenenergies, with which we can obtain the eigenstates according to (\ref{eq:c_m}) and (\ref{eq:3recur}). 

The G-functions of the two-photon and two-mode Rabi models \cite{Duan_2015,Duan_2016,Braak_2022} have been analyzed separately, whereas few studies focused on the intensity-dependent Rabi model. In this paper, three nonlinear Rabi models can be described by a unified G-function  (\ref{eq:Gfun}), and one only need to keep in mind that they correspond to different realizations of su(1,1) algebra.

\section{Results and discussions} \label{sec:result}

\begin{figure}[htb]
	\centering
	\includegraphics[scale=0.85]{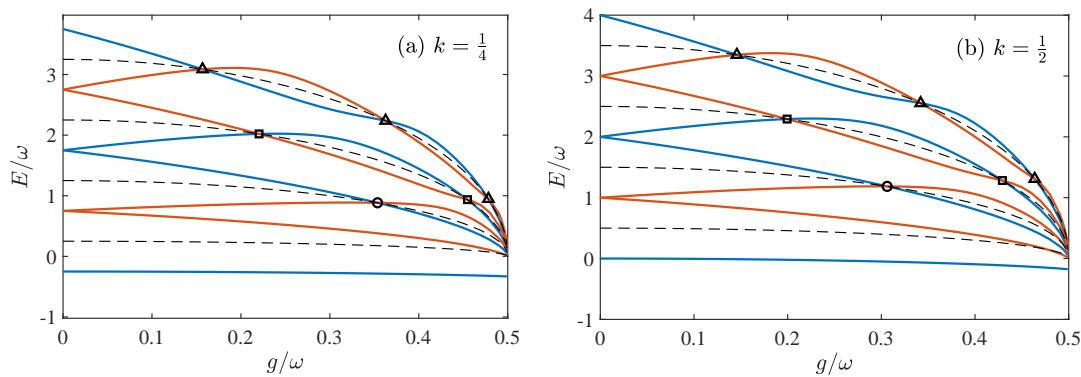} 
	\caption{Energy spectrum as a function of the coupling strength $g$ at $\epsilon = \omega = 1$ for (a) $k = \frac{1}{4}$ and (b) $k = \frac{1}{2}$. The blue (red) solid lines correspond to the even (odd) parity. The dashed lines refer to the baseline energies. Different symbols are related with the exact isolated solutions given in Table \ref{tab:Juddian}. The circle, square and triangle correspond to the exact isolated solutions with $M=1$, $2$ and $3$ respectively.}\label{fig:spectrum}
\end{figure}

The energy spectra for $k=\frac{1}{4}$ and $\frac{1}{2}$ are illustrated in figure \ref{fig:spectrum}. The two-photon Rabi model is associated with $k = \frac{1}{4}$, as shown in figure \ref{fig:spectrum}(a). It should be noted that the eigenenergies of the two-photon Rabi model should be $E_{\mathrm{2p}} = E - \frac{\omega_{\mathrm{2p}}}{2}$, as indicated in (\ref{eq:H_2P}). Figure \ref{fig:spectrum}(b) depicts the energy spectrum at $k = \frac{1}{2}$, which is associated with the two-mode and intensity-dependent Rabi models. The eigenenergies of the two-mode and intensity-dependent Rabi models should be $E_{\mathrm{2m}} = E - \omega_{\mathrm{2m}}$ and $E_{\mathrm{I}} = E - k \omega_{\mathrm{I}}$ respectively, as indicated in (\ref{eq:H_2M}) and (\ref{eq:H_I}). Especially, different symbols located at the baseline correspond to the exact isolated solutions given in Table \ref{tab:Juddian}, which can be achieved by solving (\ref{eq:Judd}). Clearly, the exact isolated solution corresponds to the level crossing between even and odd parities, and its number on each baseline is given by $M$.

\begin{figure}[htb]
	\centering
	\includegraphics[scale=0.85]{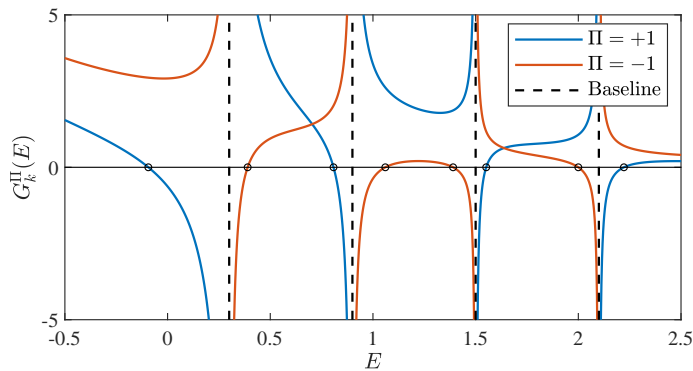} 
	\caption{$G_k^{\Pi}$ as function of $E$ for $\epsilon=\omega=1$, $g=0.4$, and $k=\frac{1}{2}$. The blue (red) solid lines correspond to the even (odd) parity. The dashed lines correspond to the baseline energies. The black circles refer to the eigenvalues obtained from the numerical diagonalization. }\label{fig:gfun}
\end{figure}

Beyond the exact isolated solutions, the regular spectrum is determined by finding the roots of the G-function. As an example, figure \ref{fig:gfun} shows the G-function at $\epsilon = \omega = 1$, $g = 0.4$ and $k = \frac{1}{4}$. It can be used to describe either the two-mode Rabi model with $\omega_{\mathrm{2m}} = \omega / 2 = 0.5$ and $g_{\mathrm{2m}} = g = 0.4$, or the intensity-dependent Rabi model with $\omega_{\mathrm{I}} = \omega = 1$ and $g_{\mathrm{I}} = g = 0.4$. The eigenenergies obtained from the numerical diagonalization are also depicted as a benchmark. The roots of $G_k^{\Pi} (E) = 0$ correspond to the eigenenergies, which fit well with those obtained from the numerical diagonalization.

\begin{figure}[htb]
	\centering
	\includegraphics[scale=0.85]{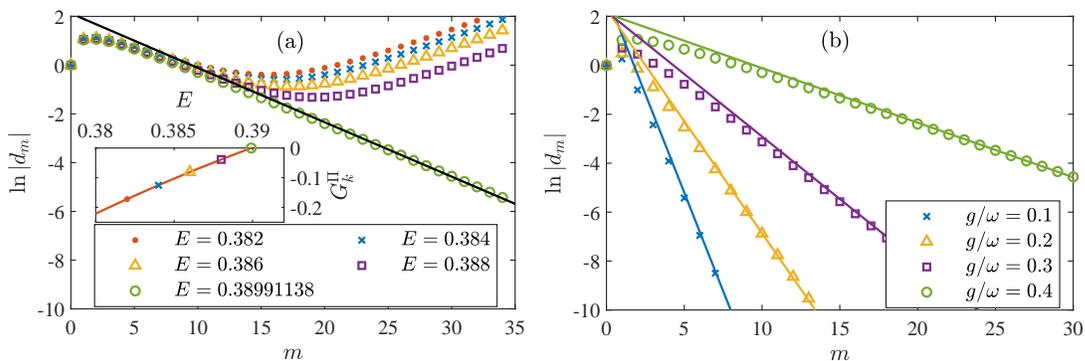} 
	\caption{$\ln|d_m|$ for $\epsilon=\omega=1$ and $k=\frac{1}{2}$. The solid lines are plotted as a benchmark, which correspond to the asymptotic behavior $|d_m| \propto \exp \left(- \gamma_{\mathrm{d}} m \right)$ with $\gamma_{\mathrm{d}}=\ln (\omega/2g)$. (a) The coupling strength is fixed at $g=0.4$. Different $E$ are marked with different symbols, and the corresponding $G_{k}^{\Pi}(E)$ are also shown in the inset. Note that $E = 0.38991138$ marked with green circle corresponds to the second black circle from left to right in figure \ref{fig:gfun}. (b) $\ln|d_m|$ for different coupling strength $g$. $E$ is fixed at the corresponding lowest eigenenergy for $\Pi=-1$. }\label{fig:d_m}
\end{figure}

Figure \ref{fig:d_m}(a) shows the expansion coefficients $d_m$ of wavefunctions near the eigenenergy, and the asymptotic behaviors are also depicted. The exact eigenenergy corresponds to the root of the G-function, which is given in the inset. As demonstrated in  (\ref{eq:limit})-(\ref{eq:general_d_m}), there exist two linearly independent solutions for $d_{m}$ at $0 < g < \omega / 2$: in the limit of $m \rightarrow +\infty$, $|d_m^{(0)}| \propto \left(2 g / \omega\right)^m = \exp \left(- \gamma_{\mathrm{d}} m \right)$, $|d_m^{(1)}| \propto \left(\omega / 2g\right)^m = \exp \left(\gamma_{\mathrm{d}} m \right)$, with the decay rate $\gamma_{\mathrm{d}}=\ln (\omega/2g)$.  The decay rate $\gamma_{\mathrm{d}}$ only depends on $g$ and $\omega$ rather than $k$. Due to $\gamma_{\mathrm{d}}>0$, $d_m^{(0)}$ tends to exponentially decay, whereas $|d_m^{(1)}|$ tends to exponentially increase. As indicated by the red dots in figure \ref{fig:d_m}(a), when $E$ is far away from the eigenenergy, $d_m$ obtained from (\ref{eq:3recur}) is dominated by $d_m^{(1)}$. When $E$ is closer to the eigenenergy, the weight of $d_m^{(0)}$ increases. For small $m$, $d_m$ shows an exponential decay behavior which is governed by $d_m^{(0)}$. On the contrary,  an exponential increase behavior emerges for large $m$ which is governed by $d_m^{(1)}$. When $E$ is exactly the eigenenergy, we expect that $d_m = d_m^{(0)}$ which leads to exponential decay of the expansion coefficients of the wavefunction, as illustrated by the green circles in figure \ref{fig:d_m}(a).

The influences of the coupling strength $g$ on $d_m$ are shown in figure \ref{fig:d_m}(b), where we have chosen $E$ the exact eigenenergies. The asymptotic behavior in the limit of $m \rightarrow +\infty$ is well described by  $|d_m^{(0)}| \propto \exp \left(- \gamma_{\mathrm{d}} m \right)$. As indicated by the blue crossing, the expansion coefficients decay very fast for weak coupling strength, and one can easily obtain the convergent eigenstates. Increasing the coupling strength $g/\omega$ leads to the decease of $\gamma_{\mathrm{d}}$, which indicates that one need more bases to describe the corresponding eigenstate. When the coupling strength tends to the spectral collapse point $g \rightarrow \omega / 2$, the decay rate $\gamma_{\mathrm{d}}$ tends to zero. Therefore, one can hardly describe the properties near the spectral collapse point with a truncated Hilbert space.

\section{Summary} \label{sec:summary}

In the last decades, exploring the strong and nonlinear coupling between light and matter has achieved great processes. The interest in the nonlinear Rabi models has blossomed both experimentally and theoretically. In this paper, we focus on three typical nonlinear Rabi models: two-photon, two-mode and intensity-dependent Rabi models, and propose a unified analytical approach. 

Previous studies mainly dealt with three models individually, and their common behaviors didn't receive sufficient attention. By virtue of different realizations of the su(1,1) Lie algebra, three models can be described by the same Hamiltonian with $\mathcal{Z}_2$ symmetry. By choosing appropriate basis states, we construct an ansatz to describe the eigenstates, whose expansion coefficients satisfy a three-term recurrence relation. 
Of special significance is the baseline energy identifying the exact isolated solutions at the level crossings between different parities, for which the eigenstates can be reduced to a closed form.
We reproduce the exact isolated solutions in the two-photon and two-mode Rabi models, whereas those in the intensity-dependent Rabi models are first achieved.

Beyond the exact isolated solutions, we propose a unified G-function based on the $\mathcal{Z}_2$ symmetry, whose roots give the regular spectrum. The expansion coefficients of the eigenstates present an exponential decay behavior in the limit of $m \rightarrow +\infty$, and the decay rate $\gamma_{\mathrm{d}}$ can be achieved analytically. With increasing coupling strength $g/\omega$, the decay rate $\gamma_{\mathrm{d}}$ decreases and it tends to zero in the spectral collapse point $g \rightarrow \omega / 2$. 

With the unified analytical approach, we achieve the eigenstates and eigenenergies of three nonlinear Rabi models, and their common behaviors are addressed. The nonlinear Rabi model introduces new physical mechanisms which cannot be captured by the linear one. The squeezing effect is inherent in the nonlinear Rabi model, compared to the linear one which is obvious only if the frequency of the two-level system is large enough \cite{PhysRevA.94.063824}. The exotic nonlinear phenomena, the squeezing effect, and their applications in quantum information deserve further consideration, which are left to future research.

\ack
We wish to thank Daniel Braak and Qing-Hu Chen for stimulating discussions.

\section*{References}
\bibliographystyle{unsrt}
\bibliography{ref}

\end{document}